\documentclass[%
preprint,
showpacs,preprintnumbers,
 amsmath,amssymb,
]{revtex4-1}

\usepackage{threeparttable}
\usepackage{graphicx}% Include figure files
\usepackage{dcolumn}% Align table columns on decimal point
\usepackage{bm}% bold math
\usepackage{hyperref}% add hypertext capabilities
\begin{document}

%\preprint{APS/123-QED}

\title{Solution of Two-Body Bound State Problems with Confining Potentials}

\author{M. R. Hadizadeh$^1$}
\email{hadizade@ift.unesp.br}

\author{Lauro Tomio$^1$,$^2$}

\affiliation{$^1$Instituto de F\'{\i}sica Te\'{o}rica (IFT), Universidade Estadual Paulista (UNESP), Barra Funda, 01140-070, S\~{a}o Paulo, Brazil,}
\affiliation{$^2$Instituto de F\'{\i}sica, Universidade Federal Fluminense, 24210-346, Niter\'oi, RJ, Brazil,}

\date{\today}% It is always \today, today,

\begin{abstract}
The homogeneous Lippmann-Schwinger integral equation is solved in momentum space by using confining potentials. Since the confining potentials are unbounded at large distances, they lead to a singularity at small momentum. In order to remove the singularity of the kernel of the integral equation, a regularized form of the potentials is used. As an application of the method, the mass spectra of heavy quarkonia, mesons consisting from heavy quark and antiquark $(\Upsilon(b\bar{b}), \psi(c\bar{c}))$, are calculated for linear and quadratic confining potentials. The results are in good agreement with configuration space and experimental results.
\end{abstract}

%Use showkeys class option if keyword
                              
\maketitle

\section{Numerical solution of non-relativistic equation }

The solution of non-relativistic and semi-relativistic Schr\"{o}dinger equation with confining potentials is interesting in various phenomena in physics, from particle to atomic physics. Many numerical methods have been developed to study such systems in configuration and momentum spaces, such as recent asymptotic exact solution of two- and three-body problems \cite{Day-FBS47,McEwen-FBS47}. In this work we use a regularization method to study the two-body systems, which interact by confining potentials. The bound state of two equal mass particles in momentum space and in a PW representation is described as:
\begin{eqnarray}
\label{eq.LS}
\psi_{l} (p) &=& \frac{1}{E-\frac{p^2}{m}} \, \int_{0} ^{\infty} dp' \, p'^2 \, V_l (p,p') \,
\psi_{l} (p'), \\ \label{eq.Vl}
V_l (p,p') &=& 2 \pi \int_{-1} ^{+1} dx \, P_l(x) \, V(p,p',x).
\end{eqnarray}
Since the integral equation (\ref{eq.LS}) is singular for confining potentials, consequently the calculated energy eigenvalues would not be in agreement with the exact analytic binding energies. To overcome this problem one can use the regularized form of confining potentials to remove the singularity of the kernel. To this aim one can keep the divergent part of the potential fixed after exceeding a certain distance, which creates an artificial barrier. The influence of tunneling barrier is manifested by significant changes in the energy eigenvalues at small distances. For numerical solution of integral equations (\ref{eq.LS}) and (2) and for discretization of continuous momentum and angle variables we have used Gauss-Legendre quadrature grids with hyperbolic plus linear (200 mesh points) and linear (100 mesh points) mapping correspondingly. The momentum integration interval $[0,\infty)$ is covered by a combination of hyperbolic and linear mappings of Gauss-Legendre points from the interval $[-1,+1]$ to the intervals $\underbrace{[0,p_1] \, + [p_1,p_2]}_{hyperbolic} \, + \underbrace{[p_2,p_3]}_{linear}$ as:
 \begin{eqnarray}
 \label{eq.mapping}
P_{hyperbolic} = \frac{1+x}{\frac{1}{p_1} + (\frac{2}{p_2}-\frac{1}{p_1} )\, x }, \quad P_{linear}=\frac{p_3-p_2}{2}\, x + \frac{p_3+p_2}{2}.
\end{eqnarray}
The used values for $p_1, \, p_2$ and $p_3$ in our calculations are $1.0, \, 3.0, \, 10.0$. In the following we present the calculated energy eigenvalues for linear and quadratic harmonic oscillator potentials and we investigate the agreement to energy eigenvalues obtained from analytical solution of Schr\"{o}dinger equation.
Fourier transformation of regularized form of these potentials to momentum space is given by:
 \begin{eqnarray}
 \label{eq.Linear+Quadratic}
V(r)=a_1 r: \quad V(p,p',x) &=& a_1 r_c \delta^3 ({\bf q}) \nonumber \\  &+& \frac{a_1}{2 \, \pi^2 \, q^4} \biggl
(2 \cos(q \, r_c) -2 +  q \, r_c \sin(q \, r_c) \biggr ) \\
V(r)=a_2 r^2: \quad V(p,p',x) &=& a_2 r_c^2 \delta^3 ({\bf q}) \nonumber \\  &+& \frac{a_2}{\pi^2 \, q^5} \biggl (3 q \, r_c \cos(q \, r_c) + ( q^2 r_c^2 -3) \sin(q \, r_c) \biggr ) ,
\end{eqnarray}
where the potentials are kept fixed at cut-off $r_c$ and
$q = |{\bf q}|= |{\bf p}- {\bf p'}|=\sqrt{p^2+p'^2-2pp'x}$.
The PW projection of these potentials $V_l(p,p')$ can be obtained by solution of
Eq. (2).
In Tables (\ref{Table_linear}) and (\ref{Table_quadratic}) we have listed our numerical results for energy eigenvalues. The $S$-wave energy levels for linear potential are compared with corresponding configuration space results and are in excellent agreement with exact energies.
As indicated in Table (\ref{Table_quadratic}) our numerical results are in excellent agreement with corresponding exact energies $E_{n,l}=(2n+l+\frac{3}{2})\hbar \, \omega$.
\begin{table}[hbt] \label{Table_linear}
\caption {Energy eigenvalues for a linear potential. The $S$-wave ($P$ and $D$-waves) energy levels are calculated for $m=1.0$ ($1.84$), $a_1=1.0$ ($0.18263$) and $r_c=20.0$.}
\begin{threeparttable}
\begin{tabular}{cc|ccccccccccccccc}
\hline
state && $1S$ &&  $2S$ && $3S$ && $4S$ && $5S$ && $6S$ \\
\hline
$E$ && $2.3381$  && $4.0879$ && $5.5205$  && $6.7867$  && $7.9441$ && $9.0226$ \\
$E$ \cite{Faustov-IJMPA15} && $2.3373$  && $4.0865$ && $5.5190$  && $6.7814$  && $7.9514$ && $9.0119$ \\
$E=-\frac{a_1 Z_0}{(ma_1)^{\frac{1}{3}}}$\tnote{\dag} && $2.3381$ && $4.0879$ && $5.5205$ && $6.7867$ && $7.9441$ && $9.0226$ \\
\hline
state && $1P$ &&  $2P$ && $3P$ && $4P$ && $5P$ && $6P$ \\
\hline
$E$ && $0.8830$  && $1.2831$ && $1.6280$  && $1.9454$  && $2.2369$ && $2.5107$ \\
\hline
state && $1D$ &&  $2D$ && $3D$ && $4D$ && $5D$ && $6D$ \\
\hline
$E$ && $1.1160$  && $1.4789$ && $1.8044$  && $2.1041$  && $2.3844$ && $2.6496$ \\
\hline
\end{tabular}
\begin{tablenotes}
\item[\dag] $Z_0$ are zeros of the Airy function.
\end{tablenotes}
\end{threeparttable}
\end{table}
In the following we test also the accuracy of our numerical calculations for coulomb potential.
PW representation of Fourier transformation of coulomb potential $V(r)=-a_{-1}/r$ to momentum space, i.e. $V(p,p',x) = \frac{-a_{-1}}{2 \, \pi^2 \, q^2}$, can be obtained analytically as $V_l(p,p') = \frac{-a_{-1}}{\pi \, p \, p'} \, Q_l ( \frac{p^2+p'^2}{2 p \, p'} )$, where $Q_l$ is the Legendre polynomial of second kind. Clearly in the calculation of $V_l(p,p')$ one should overcome the moving singularity which appears in $Q_l$ at $p=p'$. To avoid it, one can calculate $V_l(p,p')$ by solution of integral equation (2) and by using the Gauss-Legendre quadrature integration. In Table (\ref{Table_Coulomb}) we have compared our numerical results for coulomb energy levels with corresponding exact energies. Our numerical results confirm the degeneracy of energy levels for different values of $l$.

\begin{table}[hbt] \label{Table_quadratic}
\caption {Energy eigenvalues of a quadratic potential for $m=1.0$, $a_2=0.25$ and $r_c=10.0$.}
\begin{tabular}{c|cccccccccccccccccc}
\hline
state & $1S$ &  $2S$ & $3S$ & $4S$ & $5S$ & $6S$ & $7S$ & $8S$ & $9S$\\
$E$ & $1.500$  & $3.500$ & $5.500$  & $7.500$  & $9.500$ & $11.500$ & $13.500$ & $15.500$ & $17.500$ \\ \hline
state & $1P$ &  $2P$ & $3P$ & $4P$ & $5P$ & $6P$ & $7P$ & $8P$ & $9P$\\
$E$ & $2.500$  & $4.500$ & $6.500$  & $8.500$  & $10.500$ & $12.500$ & $14.500$ & $16.500$ & $18.499$ \\ \hline
state & $1D$ &  $2D$ & $3D$ & $4D$ & $5D$ & $6D$ & $7D$ & $8D$ & $9D$\\
$E$ & $3.500$  & $5.500$ & $7.500$  & $9.500$ & $11.500$ & $13.500$ & $15.500$ & $17.500$ & $19.499$ \\
\hline
\end{tabular}
\end{table}

\begin{table}[hbt] \label{Table_Coulomb}
\caption {Coulomb energy levels for $m=1$ and $a_{-1}=1$.}
\begin{tabular}{c|ccccccccccccccc}
\hline
state && $n=1$ &&  $n=2$ && $n=3$ && $n=4$ && $n=5$ \\
\hline
$E$ && $-0.2467$  && $-0.0619$ && $-0.0278$  && $-0.1568$  && $-0.0101$ \\
$E_n=-\frac{m\,a_{-1}^2}{4n^2}$ && $-0.2500$ && $-0.0625$ && $-0.0278$ && $-0.1562$ && $-0.0100$ \\
\hline
\end{tabular}
\end{table}

\section{Heavy quarkonia Mass spectrum}

In this section we solve the integral equation (\ref{eq.LS}) to calculate the mass spectra of heavy quarkonia, mesons consisting from heavy quark and antiquark. We consider both linear and quadratic confinements. The momentum space representation of the regularized form of these potentials can be obtained as:
 \begin{eqnarray}
\label{eq.Linear-confinement}
 \hspace{-2.6cm} V(r) &=& -\frac{a_{-1}}{r}+ a_1 r + a_0: \nonumber \\  V(p,p',x) &=& \biggl (-\frac{a_{-1}}{r_c} + a_1 r_c +a_0 \biggr) \delta^3 ({\bf q})
 \nonumber \\ &&  \hspace{-0.7cm} +
  \frac{-a_{-1}}{2 \, \pi^2 \, q^2} \biggl ( 1- \frac{\sin(q \, r_c)}{q \, r_c}\biggr )
 +  \frac{a_1}{2 \, \pi^2 \, q^4} \biggl
(2 \cos(q \, r_c) -2 +  q \, r_c \sin(q \, r_c) \biggr ),  \\
 \hspace{-2.6cm}  V(r) &=& -\frac{a_{-1}}{r}+ a_2 r^2 + a_0: \nonumber \\  V(p,p',x) &=& \biggl (-\frac{a_{-1}}{r_c} + a_2 r_c^2 +a_0 \biggr) \delta^3 ({\bf q})
 \nonumber \\ &&  \hspace{-1.cm} +
  \frac{-a_{-1}}{2 \, \pi^2 \, q^2} \biggl ( 1- \frac{\sin(q \, r_c)}{q \, r_c}\biggr )
 +  \frac{a_2}{\pi^2 \, q^5} \biggl (3 q \, r_c \cos(q \, r_c) + ( q^2 r_c^2 -3) \sin(q \, r_c) \biggr ). \quad
\end{eqnarray}

In tables (\ref{Table_cc}) and (\ref{Table_bb}) the calculated Bottomonium and Charmonium mass spectra are compared with the results obtained by Faustov et al. \cite{Faustov-IJMPA15} and also experimental data \cite{Barnett-PRD54}. Our numerical results show that the regularized form of the confining potentials leads to energy eigenvalues which are in good agreement with configuration space calculations and also experimental data. The study of two-body bound states with other confining potentials and also in a relativistic frame is in progress.

\begin{table}[hbt] \label{Table_cc}
\caption {Charmonium $\psi(c\bar{c})$ mass spectrum calculated for the sum of linear and quadratic confining potentials with the coulomb potential. The parameters of calculation for linear plus coulomb (quadratic plus coulomb) potentials are as:
$a_0=-0.29 \, (-0.05) \, GeV$, $a_{-1}=\frac{4}{3}\alpha_s$; $\alpha_s=0.47 \, (0.345)$, $a_1=0.18 \, GeV^2$ ($a_2=0.174 \, GeV^3$), $r_c=10.0 \, (3.0) \, fm$ and $m_c=1.56 \, (1.55) \, GeV$. }
\begin{threeparttable}
\begin{tabular}{cccccccccccccccc}
\hline
\multicolumn{1}{c}{State} && \multicolumn{2}{c}{linear+coulomb} && \multicolumn{2}{c}{quadratic+coulomb} && \multicolumn{1}{c}{Exp. \cite{Barnett-PRD54}}\\
 \cline{3-4} \cline{6-7}
 &&  Present  & Faustov et al. \cite{Faustov-IJMPA15} &&  Present  & Faustov et al. \cite{Faustov-IJMPA15} &  \\ \hline
$1S$ && 3.062 & 3.068  && 3.076 & 3.070 && 3.0675  \\
$2S$ && 3.696 & 3.697  && 3.720 & 3.730 &&  3.663 \\
$3S$ && 4.144 & 4.144  && 4.331 & 4.331 &&  4.159\tnote{\dag}\\
$1P$ && 3.529 & 3.526  && 3.492 & 3.508 &&  3.525 \\
$2P$ && 3.997 & 3.993  && 4.108 & 4.095 &&   \\
$3P$ && 4.384 & 4.383  && 4.652 & 4.670 &&   \\
$1D$ && 3.832 & 3.829  && 3.811 & 3.841 &&  3.770\tnote{\S}\\
$2D$ && 4.237 & 4.234  && 4.396 & 4.415 && \\
\hline
\end{tabular}
\begin{tablenotes}
\item[\dag] $\, ^3S_1$ state
\item[\S] $\, ^3D_1$ state
\end{tablenotes}
\end{threeparttable}
\end{table}

\begin{table}[hbt] \label{Table_bb}
\caption {Same as table (\ref{Table_cc}) but for Bottomonium $\Upsilon(b\bar{b})$ mass spectrum. $m_b=4.93 \, (4.95) \, GeV$ and $\alpha_s=0.39$ for linear plus coulomb potentials. Other potential parameters are the same as previous table.}
\begin{threeparttable}
\begin{tabular}{cccccccccccccccc}
\hline
\multicolumn{1}{c}{State} && \multicolumn{2}{c}{linear+coulomb} && \multicolumn{2}{c}{quadratic+coulomb} && \multicolumn{1}{c}{Exp. \cite{Barnett-PRD54}}\\
 \cline{3-4} \cline{6-7}
 &&  Present  & Faustov et al. \cite{Faustov-IJMPA15} &&  Present  & Faustov et al. \cite{Faustov-IJMPA15} &  \\ \hline
$1S$ && 9.425 & 9.447  &&  9.730  & 9.447 && 9.4604\tnote{\dag}\\
$2S$ && 10.006 & 10.012 && 10.014 & 10.007 && 10.023\tnote{\dag}\\
$3S$ && 10.350 & 10.353 && 10.379 & 10.389 && 10.355\tnote{\dag}\\
$4S$ && 10.628 & 10.629 && 10.724 & 10.742 && 10.580\tnote{\dag}\\
$1P$ && 9.909 & 9.900   && 9.892  & 9.898  && 9.900 \\
$2P$ && 10.263 & 10.260 && 10.265 & 10.259 && 10.260 \\
$3P$ && 10.546 & 10.544 && 10.594 & 10.593 && \\
$1D$ && 10.158 & 10.155 && 10.135 & 10.147 &&  \\
$2D$ && 10.450 & 10.448 && 10.488 & 10.486 &&  \\
\hline
\end{tabular}
\begin{tablenotes}
\item[\dag] $\, ^3S_1$ state
\end{tablenotes}
\end{threeparttable}
\end{table}

\section*{acknowledgments}
We would like to thank the Brazilian agencies FAPESP and CNPq for partial support.

\end{document}